\documentclass[twocolumn,showpacs,%
  nofootinbib,aps,superscriptaddress,%
  eqsecnum,prd,notitlepage,showkeys,10pt]{revtex4-1}

\usepackage{amssymb}
\usepackage{amsmath}
\usepackage{graphicx}
\usepackage{dcolumn}
\usepackage{hyperref}
\usepackage{multirow}

\begin{document}

\title{Core electron binding energies of adsorbates on Cu(111) from first-principles calculations}
\author{J. Matthias Kahk}
\affiliation{Department of Materials, Imperial College London, South Kensington, London SW7 2AZ, United Kingdom}
\author{Johannes Lischner}
\affiliation{Department of Physics and Department of Materials, and the Thomas Young Centre for Theory and Simulation of Materials, Imperial College London, London SW7 2AZ, United Kingdom}

\begin{abstract}
Core-level X-ray Photoelectron Spectroscopy (XPS) is often used to study the surfaces of heterogeneous copper-based catalysts, but the interpretation of measured spectra, in particular the assignment of peaks to adsorbed species, can be extremely challenging. In this study we demonstrate that first principles calculations using the delta Self Consistent Field (delta-SCF) method can be used to guide the analysis of experimental core level spectra of complex surfaces relevant to heterogeneous catalysis. Specifically, we calculate core-level binding energy shifts for a series of adsorbates on Cu(111) and show that the resulting C1s and O1s binding energy shifts for adsorbed CO, CO$_2$, C$_2$H$_4$, HCOO, CH$_3$O, H$_2$O, OH, and a surface oxide on Cu(111) are in good overall agreement with the experimental literature. In the few cases where the agreement is less good, the theoretical results may indicate the need to re-examine experimental peak assignments.

\end{abstract}

\maketitle

\section{Introduction}

Metallic copper and copper nanoparticles play an important role in industrially relevant catalytic processes, such as the low-temperature water gas shift reaction \cite{Newsome80,Zhang17,Nielsen18} and the synthesis of methanol from CO$_{2}$ and H$_{2}$ \cite{Kattel16,Behrens14,Meunier11}. Considerable efforts have been directed towards understanding the mechanisms that operate in these systems and X-ray photoemission spectroscopy (XPS) has been the tool of choice in many experimental studies \cite{Eren15,Deng08,Mudiyanselage13,Graciani14,Rodriguez07,Favaro17,Nakamura97,Nishimura00,Pohl98,Stacchiola15,Carley94}. XPS is particularly attractive for the characterization of surfaces because it provides information about the elemental composition of the surface as well as the chemical states of the elements. 

However, despite nearly forty years of research, many gaps remain in our understanding of the correspondence between features in the experimental XPS spectra and the composition of the sample surface. For example, O1s peaks at binding energies of 531.4 eV, 533.4 eV, 534.2 eV and 535.5 eV have been assigned to physisorbed CO$_{2}$ on Cu(111), polycrystalline Cu, Cu(211) and Cu(100), respectively \cite{Favaro17,Shuai13,Copperthwaite88,Browne91}. It is surprising that the reported values differ by as much as 4 eV as physisorbed CO$_2$ is expected to interact only weakly with any of these surfaces. The situation is similar for many other species: for HCOO$^-$ (formate) on Cu(111), C1s binding energies ranging from 287.3 eV to 289.8 eV have been reported \cite{Carley96,Nakamura97,Nishimura00,Pohl98}, and values between 288.2 eV and 291.0 eV have been assigned to the C1s peak of ``surface carbonates" on various copper surfaces \cite{Carley96,Deng08,Browne91}. Importantly, the reported binding energy ranges for these species also overlap with reported binding energies of ``chemisorbed CO$_2$" which range from 287.9 eV to 289.8 eV \cite{Copperthwaite88,Deng08,Mudiyanselage13,Favaro17,Pohl98}. This clearly shows that there is a need for additional insights to analyze and interpret experimental photoemission spectra of adsorbed species on Cu surfaces. 

First-principles calculations based on density-functional theory (DFT) or the GW approach are routinely used to guide the interpretation of valence electron photoemission spectra \cite{Minar2011,Haverkort08,Jin13,Payne11,Kahk14,Lischner15}. In contrast, the vast majority of experimental core-level photoemission spectra are currently interpreted without the aid of computational simulation of the spectroscopic process. For example, none of the twenty experimental XPS studies of Cu surfaces that we reviewed when writing the manuscript used comparisons to theoretical core level binding energies to guide peak fitting \cite{Favaro17,Bowker80,Copperthwaite88,Browne91,Carley96,Pohl98,Nishimura00,Rodriguez07,Deng08,Yang10,Shuai13,Mudiyanselage13,Stacchiola15,Hofmann94,Nakamura97,Au80,Eren15,Graciani14,Roberts14,Carley94}. However, a number of approaches for calculating core level bindings energies have been developed over the years, including the frozen-orbital method \cite{Zhao17}, the Z+1 approximation \cite{Delesma18}, the Slater-Janak transition state method \cite{Kunkel18,Artyushkova17,Bellafont17,Artyushkova13,Aizawa05}, the GW method \cite{vanSetten18} and the $\Delta$-SCF scheme \cite{Cavigliasso99,Cavigliasso99-1,Bellafont2016}. In the $\Delta$-SCF scheme, the core-level binding energy is calculated as the total energy difference between the ground state and the fully screened final state. Benchmark calculations on molecular systems indicate that $\Delta$-SCF calculations based on DFT yield binding energies shifts within 0.3~eV of the experimental values \cite{Cavigliasso99,Cavigliasso99-1,Bellafont2016}. This accuracy is significantly higher than reported binding energy ranges for many adsorbates on Cu surfaces and therefore insights from theoretical calculations should be very useful for the interpretation of experimental core level spectra.

In this paper we use the $\Delta$-SCF method to calculate core-electron binding energies of various adsorbed species on Cu(111), which is the lowest energy surface of metallic copper. In particular, we determine C1s and O1s binding energy shifts for CO, CO$_2$, ethene, formate, methoxy, water, OH, and a surface oxide on Cu(111). We compare our calculations in detail with the available experimental literature and highlight cases where experimental peak assignments need to be re-examined. 

\section{Computational details}
The calculations of core-electron binding energies of adsorbed molecules on Cu(111) were performed in two stages, as described below.

\begin{figure}
	\centering
	\includegraphics[width=3.33in]{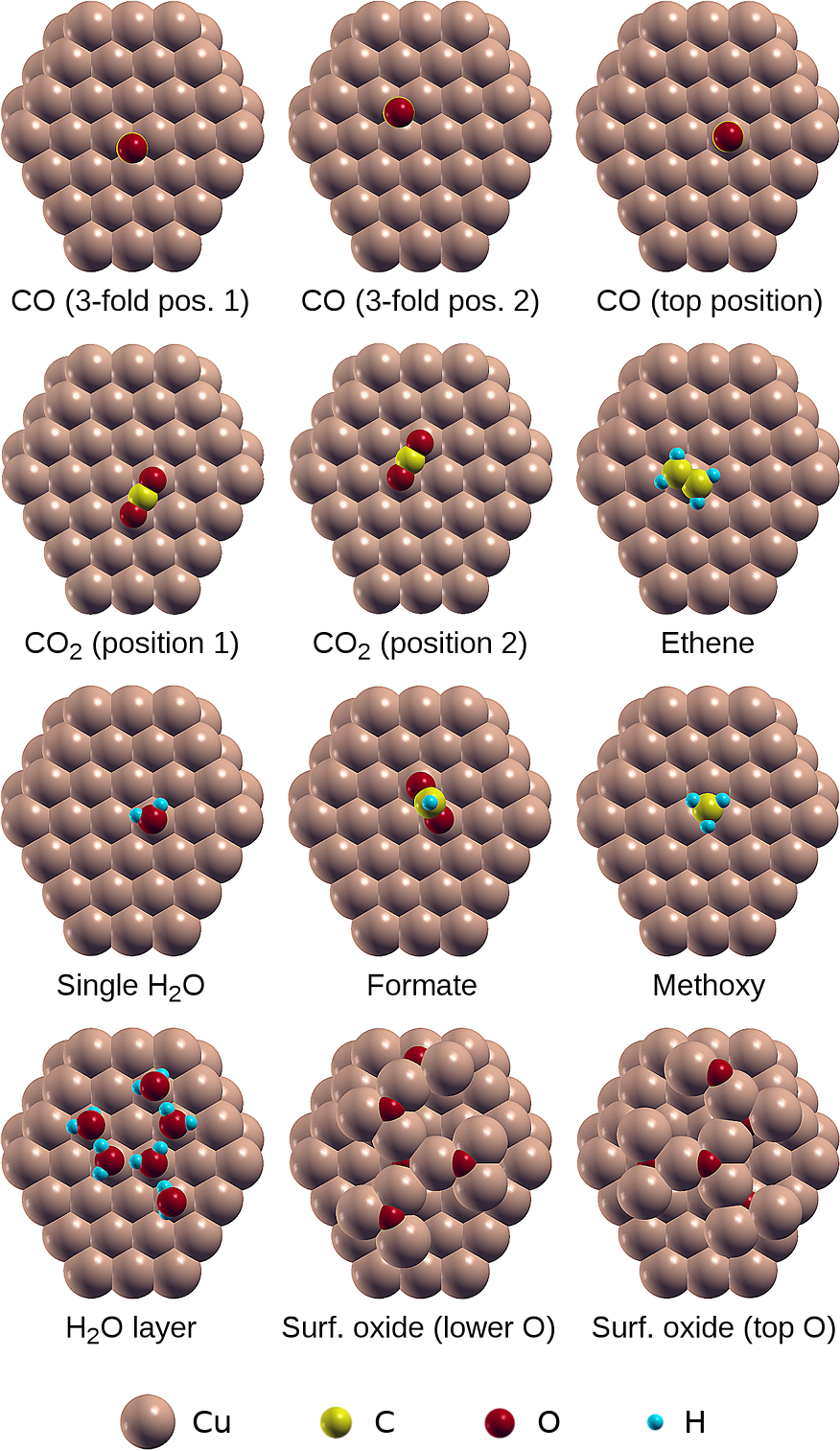}
	\caption{The clusters used for the calculation of core level binding energies of adsorbates on Cu(111).}
	\label{Fig_clustergeom}
\end{figure}

Firstly, adsorption geometries for all of the adsorbed species in the ground state were obtained by using a slab model of the surface. To generate starting configurations, the results of previously published experimental and theoretical studies on the adsorption of CO \cite{Zhang14,Gajdoš05}, CO$_2$ \cite{Wang07,Muttaqien17_2,Muttaqien17}, C$_2$H$_4$ \cite{Watson00}, H$_2$O \cite{Phatak09,Michaelides06,Michaelides07,Carrasco12}, HCOO$^-$ \cite{Nakamura97,Yang10}, CH$_3$O$^-$ \cite{Greeley02,Chen06,Hofmann94} and OH$^-$ \cite{Phatak09} on Cu(111), as well as the study of Lian et al. on the formation of surface oxides on low-index Cu surfaces \cite{Lian16}, were used. For the case of adsorbed water, we have considered two distinct models: an isolated H$_2$O molecule on Cu(111) and an H$_2$O molecule hydrogen bonded to two other surface H$_2$O molecules, with a similar local environment to what is found in water hexamers on Cu(111) \cite{Michaelides07,Carrasco12}. For the case of adsorbed CO, we have considered two distinct adsorption sites: the ``top" site, directly above a surface Cu atom, and the ``three-fold" site, in the valley between three surface Cu atoms (see Figure \ref{Fig_clustergeom} and Supplementary Figure 2). The Cu slabs were cut with the (111) faces exposed and are four atomic layers thick. In order to minimize the interactions between periodic images, the slabs were built from orthorhombic supercells with a total of 64 Cu atoms per cell, except for the case of the surface oxide for which a 4 $\times$ 4 supercell of the hexagonal Cu(111) surface unit cell was used.

The structures were relaxed until the forces on the atoms were less than $10^{-3}$ Ry/bohr and the total energy change between the last two optimization steps was less than $10^{-4}$ Ry. These calculations were carried out using DFT as implemented in the Quantum Espresso software package \cite{Giazonni09}, which employs a plane-wave basis set. Cut-off energies of 40 Ry and 200 Ry were used for the wavefunctions and the charge density, respectively, and the interaction between core and valence electrons is described via ultrasoft pseudopotentials from the Garrity-Bennett-Rabe-Vanderbilt (GBRV) Pseudopotential Library \cite{Garrity14}. The slabs were separated by $\sim 14$~{\AA} of vacuum and a dipole correction \cite{Bengtsson99} was used to minimize spurious interactions between adjacent layers. We employed the PBE exchange-correlation functional \cite{Perdew96} with the Grimme-D2 correction to capture the effect of van der Waals interactions \cite{Grimme06}. The relaxed geometries are shown in Supplementary Figure 2 and the corresponding atomic positions are also provided in the supplementary materials.\dag

Secondly, photoelectron binding energies were calculated using the $\Delta$-SCF approach, i.e. as the total energy difference between the ground state and the ionized state where one electron is removed from a core orbital. This corresponds to the assumption of a fully screened core hole. In these calculations, the surfaces were modelled as clusters cut from the relaxed slabs generated in the first step. The clusters, comprising 88 Cu atoms and the adsorbate, were cut such that the adsorbed species sits approximately at the centre of the top (111) face. The geometries of the adsorbed species on the clusters are illustrated in Figure~\ref{Fig_clustergeom} and the corresponding atomic positions are provided in supplementary information.\dag 

The total energies of the clusters were calculated using DFT with a Gaussian orbital basis set as implemented in the all-electron quantum chemistry code NWChem \cite{Valiev10}. For simulating the final states, an explicit core hole was generated by constraining the occupancy of one of the core orbitals, whilst all other electrons were allowed to relax in the presence of the core hole. The basis sets used in the cluster calculations are provided in the supplementary information.\dag Briefly, effective core potentials with the associated basis sets from reference \cite{Stevens92} were used for the Cu atoms with the following modifications: for the Cu atoms in the top layer, the exponents of the two most diffuse sp-type basis functions were increased to 0.1619 and 0.074, respectively; for the Cu atoms which are not in the top layer, the sp-type basis function with the smallest exponent was removed and the exponent of the sp-type basis function with the second smallest exponent was reduced to 0.1119. These changes were required to prevent numerical instabilities during the self-consistent field procedure. The pcseg-2 all-electron basis sets developed by Jensen \cite{Jensen14} were used for the light elements H, C and O, except for the atoms with a core hole, for which a special basis set with uncontracted core orbitals was used (derived from the pcJ-3\_2006 basis sets from reference \cite{Jensen06}), in order to allow full relaxation of the other electrons on the same atom in the presence of a core hole. All $\Delta$-SCF calculations of Cu(111) clusters with adsorbates were carried out using the PBE exchange-correlation functional. 

In order to assess the accuracy of our calculations, additional C1s binding energy calculations were carried out for the free molecules CH$_4$, C$_2$H$_6$, CO, CO$_2$, CCl$_4$, and CF$_4$, and the O1s binding energy was calculated for H$_2$O, CO, CO$_2$, CH$_3$OH, and HCOOH (both O sites). In these calculations both the initial structure relaxation as well as the subsequent $\Delta$-SCF calculation were carried out using both the M06 hybrid functional \cite{Zhao08} as well as PBE \cite{Perdew96}.

We observed in our calculations that the localization of a core hole onto a single atomic site may fail when there are two or more atoms of the same element in the molecule (or cluster). For such systems, it was possible to guide the core hole to the desired site by (i) introducing a fictitious additional atomic charge of +0.1 $e$ (with $e$ being the proton charge) at that site only for the initialization of the Kohn-Sham wavefunctions, and (ii) ensuring that the basis set with uncontracted core wavefunctions that is best suited to accommodate the core hole is only used at that site.

\section{Results: tests}
To assess the accuracy of our calculations for core-level binding energies of adsorbates on Cu(111) surfaces, we have carried out test calculations of (i) core-electron binding energies of free molecules, (ii) stabilization energies of a point charge above a metallic cluster, (iii) core-electron binding energies of adsorbed small molecules at different quasi-equivalent adsorption sites on Cu clusters and (iv) the density of states (DOS) of the Cu cluster and bulk Cu metal. 

\begin{figure}
	\centering
	\includegraphics[width=3.33in]{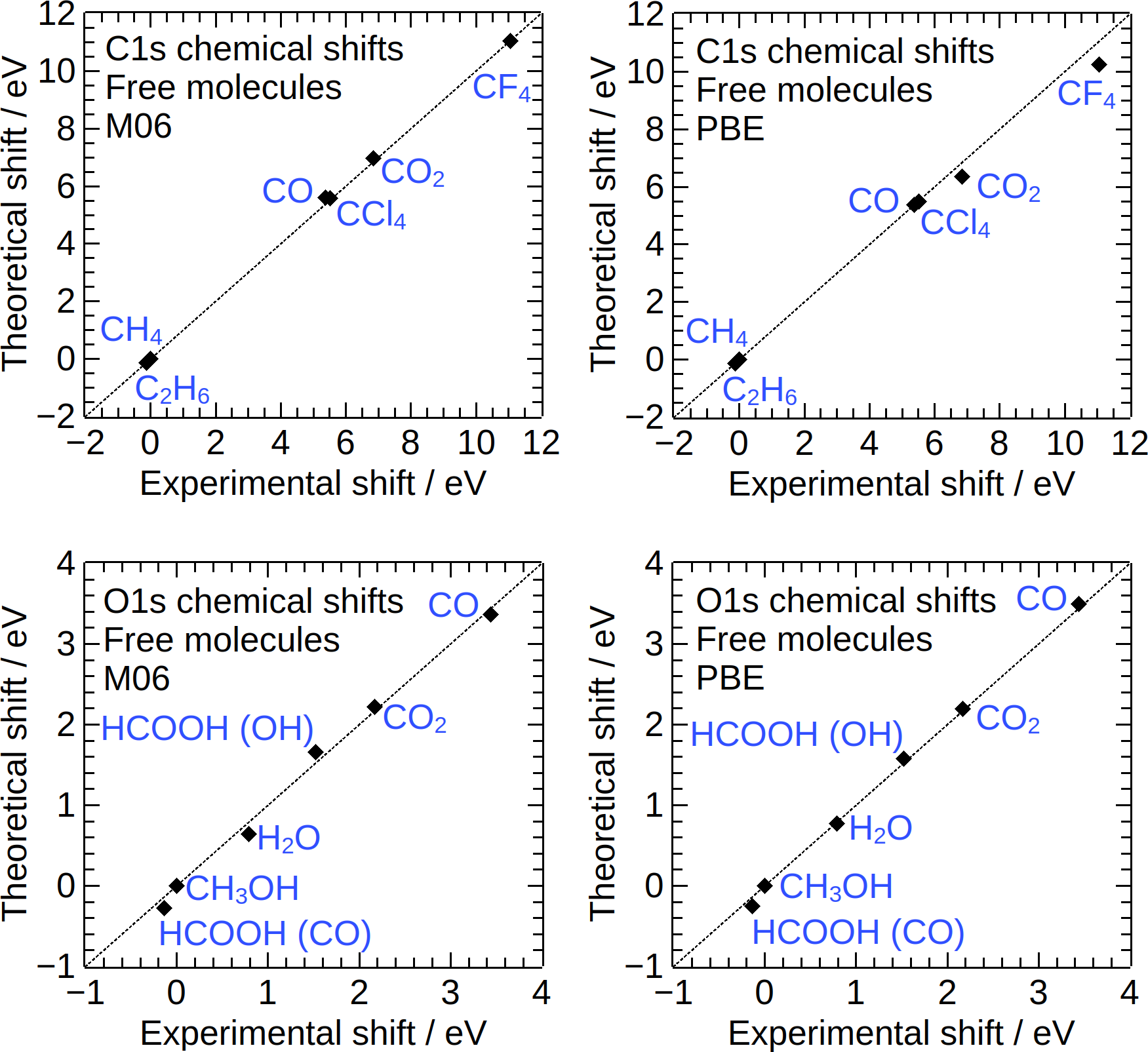}
	\caption{Theoretical core level binding energy shifts for free molecules, plotted against the corresponding experimental shifts from gas phase measurements \cite{Cavigliasso99}.}
	\label{Fig_gas_phase}
\end{figure}

The results of the calculations on free molecules are summarized in Table~\ref{Tab_gas_phase} and Figure~\ref{Fig_gas_phase}. The theoretical binding energy shifts (referenced to methane for the C1s core level and methanol for the O1s core level) have been compared to experimental values compiled by Cavigliasso \cite{Cavigliasso99}. Good agreement between theory and experiment is found for both functionals, with M06 performing somewhat better than PBE: the mean unsigned errors are 0.08 eV and 0.13 eV for M06 and PBE, respectively, and the maximum errors are 0.23 eV (C1s binding energy of CO) for M06 and 0.79 eV (C1s binding energy of CF4) for PBE. Despite the small quantitative difference with the M06 results, the results obtained with PBE are sufficiently accurate to interpret experimental spectra. It is also possible to compare the absolute values of the theoretical binding energies to the experimental data for the free molecules, and for the C1s and O1s core levels considered in this work, we find that the values agree to within $\sim 0.3$~\% for M06 and $\sim 0.5$~\% for PBE. 

\begin{table*}
\caption{A summary of the results of core-level binding energy calculations of free molecules.}
\label{Tab_gas_phase}
\begin{tabular}{ c c c c c c c}
	\hline 
	Atom\textsuperscript{\emph{a}} & Exp B.E.\textsuperscript{\emph{b}} & Exp shift & M06 shift & M06 error & PBE shift & PBE error \\
	\hline 
	\textbf{C}$_2$H$_6$ & 290.72 & -0.12 & -0.13 & -0.01 & -0.13 & -0.01 \\ 
	\textbf{C}H$_4$ & 290.84 & 0 & 0 & 0 & 0 & 0 \\
	\textbf{C}O & 296.21 & 5.37 & 5.60 & 0.23 & 5.37 & 0.00 \\
	\textbf{C}Cl$_4$ & 296.36 & 5.52 & 5.58 & 0.06 & 5.48 & -0.04 \\
	\textbf{C}O$_2$ & 297.69 & 6.85 & 6.97 & 0.12 & 6.35 & -0.50 \\
	\textbf{C}F$_4$ & 301.89 & 11.05 & 11.04 & -0.01 & 10.26 & -0.79 \\
	 & & & & & & \\	
	HC\textbf{O}(OH) & 538.97 & -0.14 & -0.28 & -0.14 & -0.25 & -0.11 \\
	CH$_3$\textbf{O}H & 539.11 & 0 & 0 & 0 & 0 & 0 \\
	H$_2$\textbf{O} & 539.90 & 0.79 & 0.64 & -0.15 & 0.77 & -0.02 \\
	HCO(\textbf{O}H) & 540.63 & 1.52 & 1.66 & 0.14 & 1.58 & 0.06 \\
	C\textbf{O}$_2$ & 541.28 & 2.17 & 2.22 & 0.05 & 2.19 & 0.02 \\
	C\textbf{O} & 542.55 & 3.44 & 3.37 & -0.07 & 3.49 & 0.05 \\	
	\hline 
\end{tabular}

\textsuperscript{\emph{a}} Bold typeface is used to indicate the position of the core hole;
\textsuperscript{\emph{b}} All experimental values are taken from reference \cite{Cavigliasso99}. All energies are given in eV.
\end{table*}

\begin{figure}
	\centering
	\includegraphics[width=3.33in]{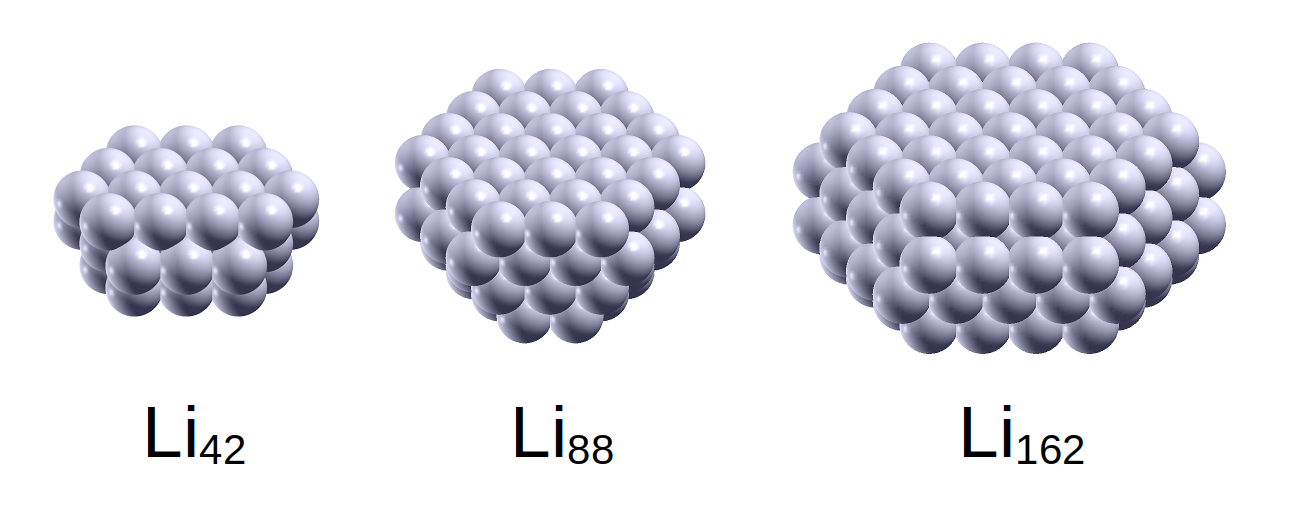}
	\caption{The clusters used for calculating the stabilization energy of a test charge near a surface of a ``model metal'' with the electronic configuration of lithium and a face-centered cubic structure.}
	\label{Fig_lithium_clusters}
\end{figure}

Next, we studied how the finite size of the cluster affects the calculated core-electron binding energies. For this, a series of calculations were performed using clusters of a ``model metal". This ``model metal" was chosen to enable the simulation of large clusters and consists of lithium atoms in a cubic close-packed structure with a lattice parameter of 4.39 {\AA}, similar to the high pressure fcc phase of lithium \cite{Smith90,Barrett47}. The shapes of the Li$_{42}$, Li$_{88}$ and Li$_{162}$ clusters used in these calculations are shown in Figure \ref{Fig_lithium_clusters}. For maximum computational efficiency, a primitive STO-2G basis set was used. A test charge of $+1~e$ located above the cluster surface was used to simulate the effect of a core hole in an adsorbed molecule. Table~\ref{Tab_Lithium} shows the calculated stabilization energies, i.e. the difference of total energies with and without the test charge, for three different cluster sizes and different heights of the test charge defined as the distance of the point charge from the plane of Cu nuclei in the top layer. The test charge was placed above either a 3-fold site or a ``top" site of the close-packed surface. We find that the stabilization energies calculated using the Li$_{88}$ cluster are within $\sim$ 0.16 eV of those obtained using the larger Li$_{162}$ cluster. If the relative stabilization energies for different sites close to the surface of the cluster are considered, the errors are even smaller. This finding suggests that the finite-size error in our calculations of adsorbates on Cu clusters consisting of 88 atoms are small enough to allow meaningful interpretation of experimental spectra. We have not been able to perform calculations on bigger Cu clusters because of the required computational expense of the calculations and the numerical instabilities that are common in simulations of metallic systems using a local orbital basis set that includes ``diffuse"  (i.e. small exponent) Gaussian basis functions. 

\begin{table*}
\caption{Stabilization energies of a test charge of +1 proton charge at various distances above the top face of a cluster of a ``model metal" with the electronic configuration of lithium. }
\label{Tab_Lithium}
\begin{tabular}{ c c c c }
	\hline 
	Position of test charge & \multicolumn{3}{c}{Stabilization energy (eV)} \\ 
	\hline 
	 & Li$_{42}$ & Li$_{88}$ & Li$_{162}$ \\ 
 	\hline
	3-fold site, $h = 1$ {\AA} & 5.91 & 5.88 & 6.04 \\
	3-fold site, $h = 2$ {\AA} & 3.79 & 3.87 & 3.98 \\
	3-fold site, $h = 3$ {\AA} & 1.99 & 2.13 & 2.17 \\
	3-fold site, $h = 5$ {\AA} & 0.69 & 0.85 & 0.87 \\
	3-fold site, $h = 10$ {\AA} & 0.12 & 0.21 & 0.23 \\
	3-fold site, $h = 30$ {\AA} & 0.01 & 0.01 & 0.01 \\
	Top site, $h = 1$ {\AA} & 4.71 & 4.78 & 4.88 \\
	Top site, $h = 3$ {\AA} & 2.04 & 2.16 & 2.20 \\
	\hline 
\end{tabular} 	
\end{table*}

We also calculated the C1s and O1s binding energies for the CO and CO$_2$ molecules adsorbed at two different adsorption sites on the top surface of the Cu$_{88}$ cluster that are equivalent with respect to the underlying lattice, but distinguishable by their position relative to the finite sized cluster, see Figure~\ref{Fig_clustergeom}. Large differences in the obtained binding energies at different quasi-equivalent sites on the cluster surface would indicate that the calculated values are strongly affected by finite size effects. The results of these tests are shown in Table~\ref{Tab_position_test}. Amongst the tested positions, the calculated binding energies vary by less than 0.05 eV indicating that finite-size effects are small.

\begin{table*}
\caption{C1s and O1s core-level binding energies of CO and CO$_2$ molecules adsorbed on different quasi-equivalent sites of a Cu$_{88}$ cluster, see Figure~\ref{Fig_clustergeom}.}
\label{Tab_position_test}
\begin{tabular}{ c c c }
	\hline
	Species & C1s theoretical B.E. (eV) & O1s theoretical B.E. (eV)\\
	\hline
	CO (3-fold pos. 1) & 289.32 & 534.81 \\
	CO (3-fold pos. 2) & 289.32 & 534.80 \\
	\multirow{2}{*}{CO$_2$ (pos. 1)} & \multirow{2}{*}{292.85} & 537.53 (O1) \\
	 & & 537.51 (O2) \\
	\multirow{2}{*}{CO$_2$ (pos. 2)} & \multirow{2}{*}{292.83} & 537.53 (O1) \\
	& & 537.52 (O2) \\
	\hline
\end{tabular}
\end{table*}

\begin{figure}
	\centering
	\includegraphics[width=3.33in]{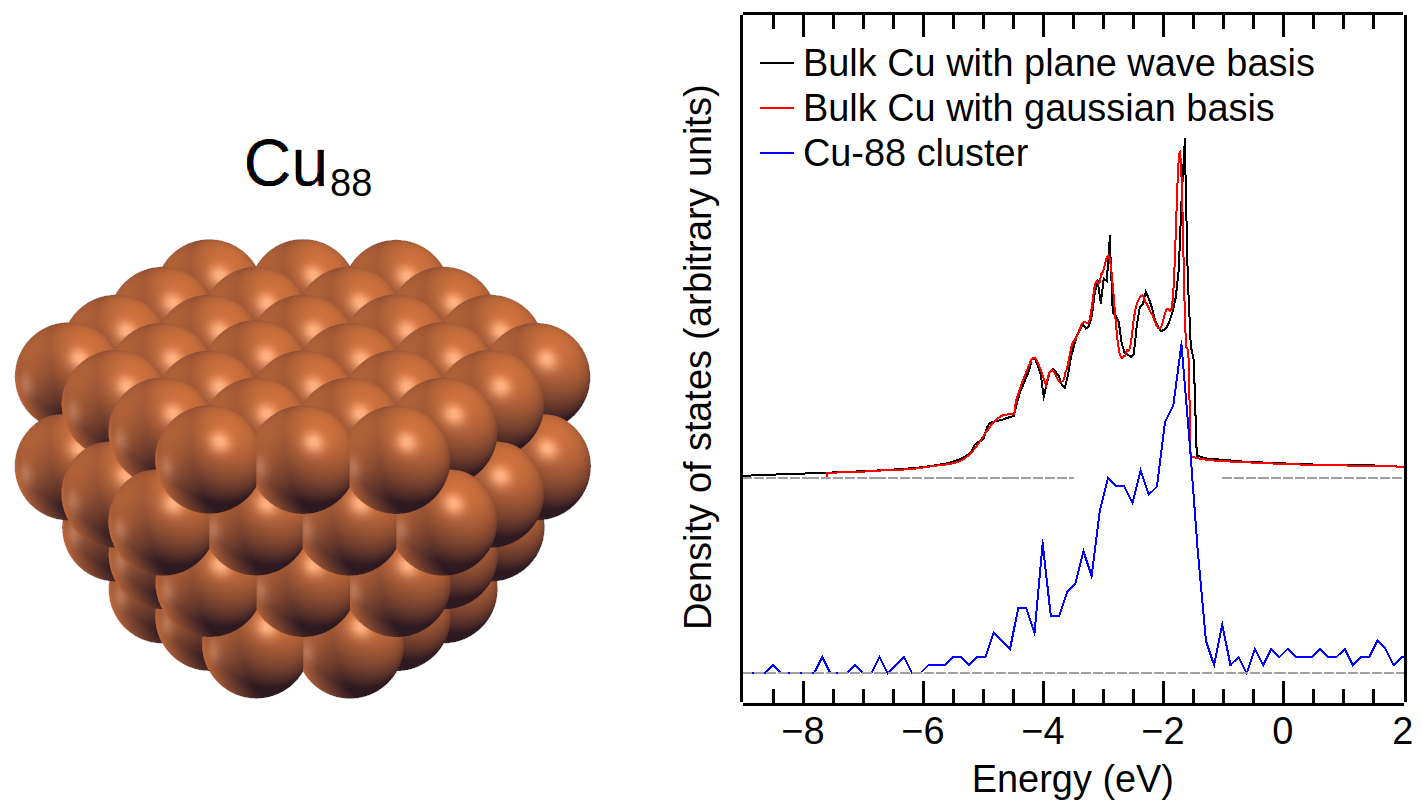}
	\caption{Left: The Cu$_{88}$ cluster used for the $\Delta$-SCF calculations. Right: The density of states of bulk Cu calculated using a plane-wave basis set, bulk Cu calculated using a Gaussian basis set and of the Cu$_{88}$ cluster. The curve of Cu$_{88}$ has been offset for clarity. The energies are referenced to the respective calculated Fermi energies.}
	\label{Fig_dos}
\end{figure}

Finally, in order to verify that the effective core potential and the Gaussian basis set from reference~\cite{Stevens92} are suitable for simulations of metallic Cu, we calculated the DOS of bulk Cu using this basis set and the CRYSTAL14 software package \cite{Dovesi14}. Figure~\ref{Fig_dos} shows that the resulting DOS is in excellent agreement with the DOS of bulk Cu obtained from plane-wave DFT. For comparison, we have also included the DOS of the bare Cu$_{88}$ cluster in Figure~\ref{Fig_dos}. 

\section{Results: Adsorbates on Cu(111)}
\begin{figure}
	\centering
	\includegraphics[width=3.33in]{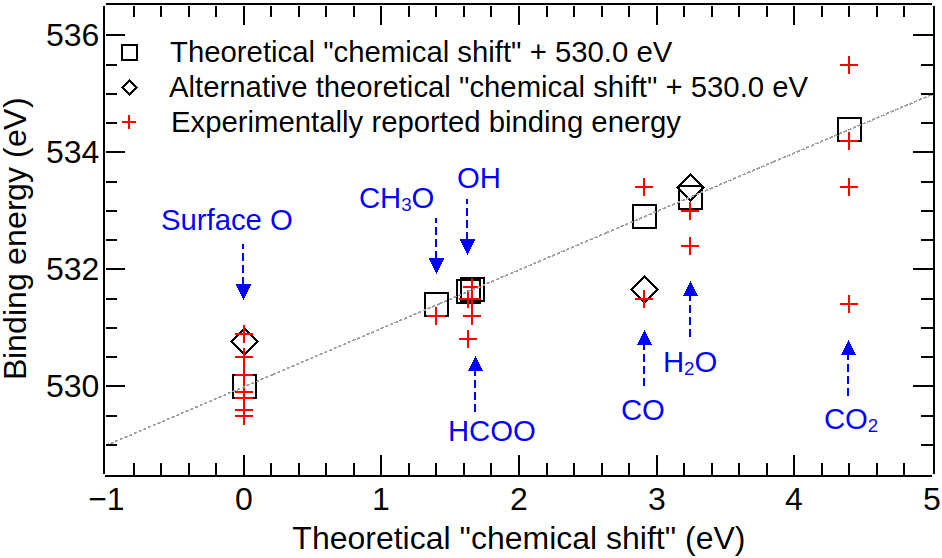}
	\caption{A comparison of calculated and experimental O1s binding energy shifts for various surface species on copper. For the surface oxide, the square indicates the theoretical value obtained for the higher (surface) oxygen site, and the diamond shows the value for the lower (buried) oxygen site. For CO, the square indicates the theoretical value obtained for the ``top" adsorption site, and the diamond shows the value for the ``3-fold" site. For water, the square indicates the theoretical value obtained for a surface water molecule hydrogen bonded to two other water molecules, and the diamond shows the value for an isolated water molecule on Cu(111). Full details and references for the experimental datapoints are given in Table \ref{Tab_O_ads}.}
	\label{Fig_O1s_shifts}
\end{figure}

\begin{figure}
	\centering
	\includegraphics[width=3.33in]{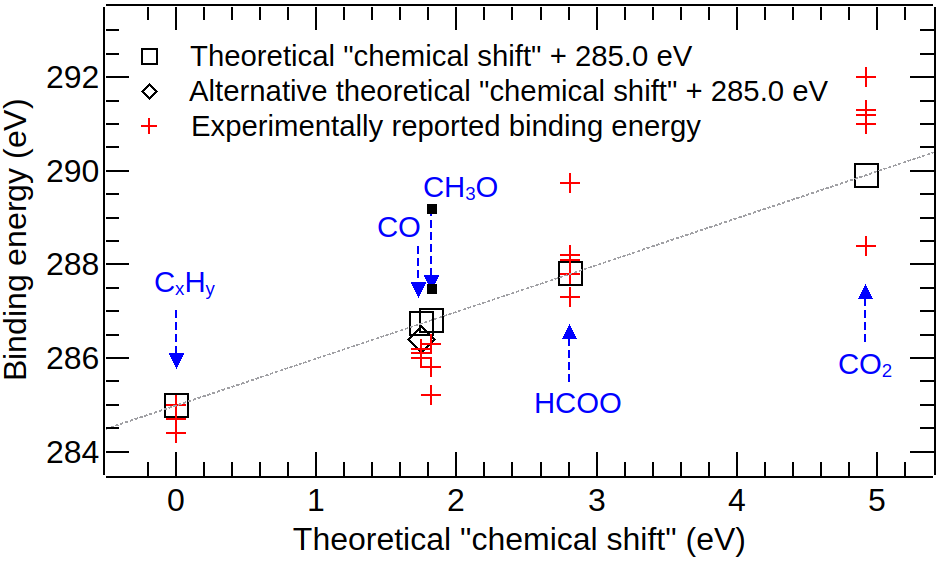}
	\caption{A comparison of calculated and experimental C1s binding energy shifts for various surface species on copper. For CO, the square indicates the theoretical value obtained for the ``top" adsorption site, and the diamond gives the value for the ``3-fold" site. Full details and references for the experimental datapoints are given in Table \ref{Tab_C_ads}.}
	\label{Fig_C1s_shifts}
\end{figure}

The results of the C1s and O1s binding energy calculations for the various adsorbed species on Cu(111) are compared to experimental data in Figures~\ref{Fig_O1s_shifts} and \ref{Fig_C1s_shifts} and in Tables~\ref{Tab_O_ads} and \ref{Tab_C_ads}. Whenever possible, we compare our results to measured binding energies of adsorbates on Cu(111). However, because of the limited availability of experimental data for this surface, we also compare to results obtained on other Cu surfaces as well as polycrystalline Cu. 

Figure~\ref{Fig_O1s_shifts} and Table~\ref{Tab_O_ads} show good overall agreement between the calculated O1s binding energy shifts and experimental measurements. In our calculations, the top O atoms in the surface oxide structure, see Figure~\ref{Fig_clustergeom}, exhibit the smallest binding energy and we use this energy as reference for all O1s binding energy shifts. The binding energy shift of the lower O atom in the surface oxide is 0.78~eV. To compare this result to experimental data, we have grouped together all peak assignments that are referred to as ``adsorbed oxygen", ``oxygen adatom" or ``surface oxide" in the experimental literature, see Table~\ref{Tab_O_ads}. The corresponding experimental binding energy shifts are calculated relative to a reference energy of 530.0~eV and range from -0.50 to +1.00~eV \cite{Favaro17,Roberts14,Stacchiola15,Copperthwaite88,Carley96,Deng08}. For an adsorbed methoxy (CH$_3$O) group we obtain a binding energy shift of 1.40~eV in good agreement with the experimental result of 1.20~eV for methoxy on Cu(110) \cite{Carley96}. The calculated binding energy shifts of adsorbed hydroxyl (OH) and formate (HCOO) are 1.63~eV and 1.66~eV, respectively, in very good agreement with the measured values of 1.50~eV for OH on Cu(111) \cite{Mudiyanselage13} and 1.50~eV for formate on Cu(111) \cite{Nakamura97,Nishimura00}. 

For the case of CO on Cu(111), it is important to note that the top adsorption site is found to be the most favourable one by experiment \cite{Raval88} and also in calculations using the DFT+U method \cite{Gajdoš05}, hybrid functionals \cite{Stroppa07} and the Random Phase Approximation (RPA) \cite{Ren09}. In contrast, standard functionals based on the Generalized Gradient Approximation (GGA) predict adsorption at the 3-fold site to be most stable \cite{Lopez01,Gajdos04}. In our calculations, a core level binding energy shift of 2.91~eV is obtained for the molecule on the top site, whereas a value of 1.67~eV is obtained for the molecule at the three-fold site. The value obtained for the top site is in reasonable agreement with the experimental value of 3.40~eV reported in reference \cite{Mudiyanselage13} for CO on the Cu(111) surface. In contrast, a much smaller binding energy shift of 1.5~eV has been reported in reference \cite{Eren15}, also for CO on Cu(111), and this value is similar to our calculated result for the three-fold site. Whilst this result might be interpreted to mean that both adsorption sites can be occupied under the measurement conditions, further work on this matter is desirable because to the aforementioned limitations of GGA functionals which we employ in our $\Delta$-SCF calculations in describing the adsorption of CO on Cu(111). 

For H$_2$O on Cu(111), both the isolated molecule and the monolayer have similar binding energy shifts of 3.41~eV and 3.24~eV, respectively. Both of these values are in good agreement with the experimental studies that report a binding energy shift of 3.0~eV for adsorbed water on Cu \cite{Roberts14,Deng08}. Finally, CO$_2$ on Cu(111) exhibits the largest binding energy shift of 4.39~eV of all oxygen-containing molecules in our study. We find that this molecule is not chemically bonded to the surface. Experimental findings for O1s binding energies of physisorbed CO$_2$ on Cu surfaces range from 1.40~eV to 5.50~eV \cite{Favaro17,Shuai13,Copperthwaite88,Browne91}, making it difficult to assess the agreement between theory and experiment. It is interesting to compare our results also to experimental measurements for physisorbed CO$_2$ on different metals as the binding energy shifts which are dominated by electrostatic image charge effects should only weakly depend on the chemical composition of the metal. In particular, O1s binding energy shifts of 4.7~eV and 5.0~eV have been reported for physisorbed CO$_2$ on Ni(110) and polycrystalline Fe, respectively \cite{Illing88}. This, combined with the theoretical results, suggests that the peaks at much lower binding energies that have been assigned to physisorbed CO$_2$ on Cu may actually correspond to some other chemical environments.

\begin{table*}
\caption{A summary of the results of O1s binding energy calculations of various adsorbates on Cu(111), as well as the surface oxide.}
\label{Tab_O_ads}
\begin{tabular}{ c c c c c c }
	\hline
	Exp. species & Ref. & Exp. B.E.\textsuperscript{\emph{a}} & Exp. shift & Theor. species & Theor. shift \\
	\hline
	Subsurface O on Cu(111) & \cite{Favaro17} & 529.80 & -0.20 & \multirow{10}{*}{Surf-ox./Cu$_{88}$} & \\
	Oxide on Cu(111) & \cite{Roberts14} & 530.50 & 0.50 & & \\
	Surface O on Cu(111) & \cite{Stacchiola15} & 530.90 & 0.90 & & \\
	Surface O on Cu(111) & \cite{Favaro17} & 531.00 & 1.00 & & 0.0 \\
	Chemisorbed oxygen on Cu(211) & \cite{Copperthwaite88} & 529.50 & -0.50 & & (top O) \\
	Adsorbed oxygen on Cu(110) & \cite{Carley96} & 529.60 & -0.40 & & 0.78 \\
	Surface O on suboxidic Cu$_x$O & \cite{Favaro17} & 529.60 & -0.40 & & (lower O) \\
	Chemisorbed O on Cu(poly) & \cite{Deng08} & 529.80 & -0.20 & & \\
	Cu$_2$O & \cite{Stacchiola15} & 529.90 & -0.10 & & \\
	Cu$_2$O & \cite{Deng08} & 530.20 & 0.20 & & \\
 	 & & & & & \\
	Methoxy on Cu(110) & \cite{Carley96} & 531.20 & 1.20 & CH$_3$O/Cu$_{88}$ & 1.40 \\
	& & & & & \\
	OH on Cu(111) & \cite{Mudiyanselage13} & 531.50 & 1.50 & \multirow{2}{*}{OH/Cu$_{88}$} & \multirow{2}{*}{1.63} \\
	OH on Cu(poly) & \cite{Deng08} & 530.80 & 0.80 & & \\
	 & & & & & \\
	Formate on Cu(111) & \cite{Nakamura97,Nishimura00} & 531.5 & 1.50 & \multirow{3}{*}{HCOO/Cu$_{88}$} & \multirow{3}{*}{1.66} \\
	Formate on Cu(110) & \cite{Carley96} & 531.20 & 1.20 & &\\
	HCOO- on cold deposited Cu film	 & \cite{Pohl98} & 531.7 & 1.70 & & \\
    & & & & & \\
	 & & & & \multirow{4}{*}{CO/Cu$_{88}$} & 1.67 \\
	 CO on Cu(111) & \cite{Eren15} & 531.50 & 1.50 & & (3-fold site) \\
	 CO on Cu(111) & \cite{Mudiyanselage13} & 533.40 & 3.40 & & 2.91 \\
	 & & & & & (top site) \\
	 & & & & & \\
	Adsorbed H$_2$O on Cu(111) & \cite{Roberts14} & 533.00 & 3.00 & & \\
	H$_2$O on Cu(111) & \cite{Favaro17} & 532.40 & 2.40 & H$_2$O/Cu$_{88}$ & 3.41 \\
	H$_2$O on Cu(poly) & \cite{Deng08} & 533.00 & 3.00 & H$_2$O-layer/Cu$_{88}$ & 3.24\\
	 & & & & & \\
	Physisorbed CO$_2$ on Cu(111) & \cite{Favaro17} & 531.40 & 1.40 & \multirow{4}{*}{CO$_2$/Cu$_{88}$} & \multirow{4}{*}{4.39} \\
	Physisorbed CO$_2$ on Cu(poly) & \cite{Shuai13} & 533.40 & 3.40 & & \\
	Monolayer physisorbed CO$_2$ on Cu(211) & \cite{Copperthwaite88} & 534.20 & 4.20 & & \\
	Physisorbed CO$_2$ on Cu(100) & \cite{Browne91} & 535.50 & 5.50 & & \\
	\hline	
\end{tabular}

\textsuperscript{\emph{a}} All energies are given in eV.
\end{table*}
     
\begin{table*}
\caption{A summary of the results of C1s binding energy calculations of various adsorbates on Cu(111).}
\label{Tab_C_ads}
\begin{tabular}{ c c c c c c}
	\hline
	Exp. species & Ref. & Exp. B.E.\textsuperscript{\emph{a}} & Exp. shift & Theor. species & Theor. shift \\
	\hline
    Graphitic carbon on Cu(111) & \cite{Favaro17} & 284.50 & -0.50 & \multirow{7}{*}{C$_2$H$_4$/Cu$_{88}$} & \multirow{7}{*}{0.00} \\
    sp3 carbon on Cu(111) & \cite{Favaro17} & 285.20 & 0.20 & & \\
	C$_x$H$_y$ on Cu(111) & \cite{Mudiyanselage13} & 285.00 & 0.00 & & \\
	C$^0$ species on Cu(poly) & \cite{Deng08} & 284.40 & -0.60 & & \\
	Carbon contamination on Cu(poly) & \cite{Deng08} & 284.70 & -0.30 & & \\
	C$^0$ on Cu(poly) & \cite{Copperthwaite88} & 285.00 & 0.00 & & \\
	Graphitic carbon on Cu(100) & \cite{Browne91} & 285.00 & 0.00 & & \\
	 & & & & & 1.40 \\
    CO on Cu(111) & \cite{Eren15} & 286.10 & 1.10 & \multirow{3}{*}{CO/Cu$_{88}$} & (3-fold site) \\
	CO on Cu(111) & \cite{Mudiyanselage13} & 286.20 & 1.20 & & 1.75 \\
	Carbonyl carbon & \cite{Copperthwaite88} & 286.00 & 1.00 & & (top site) \\
	 & & & & & \\
	C-O(H) bonds on Cu(111) & \cite{Favaro17} & 286.30 & 1.30 & \multirow{3}{*}{CH$_3$O/Cu$_{88}$} & \multirow{3}{*}{1.82} \\
	Methoxy on Cu(110) & \cite{Carley96} & 285.80 & 0.80 & & \\
    Methoxy on Cu(poly) & \cite{Deng08} & 285.20 & 0.20 & & \\
	 & & & & & \\
    HCOO- on Cu(111) & \cite{Favaro17} & 287.30 & 2.30 & \multirow{6}{*}{HCOO/Cu$_{88}$} & \multirow{6}{*}{2.81} \\
	Formate on Cu(111) & \cite{Nakamura97,Nishimura00} & 288.20 & 3.20 & & \\
	HCOO on Cu(111) & \cite{Yang10} & 289.75 & 4.75 & & \\
	Formate on Cu(poly) & \cite{Deng08} & 287.30 & 2.30 & & \\
	Formate on Cu(110) & \cite{Carley96} & 287.80 & 2.80 & & \\
	HCOO- on cold deposited Cu film & \cite{Pohl98} & 288.10 & 3.10 & & \\
	 & & & & & \\
    Physisorbed CO$_2$ on Cu(111) & \cite{Favaro17} & 288.40 & 3.40 & \multirow{5}{*}{CO$_2$/Cu$_{88}$} & \multirow{5}{*}{4.92} \\ 
	Monolayer physisorbed CO$_2$ on Cu(poly) & \cite{Copperthwaite88} & 291.00 & 6.00 & & \\
	Physisorbed CO$_2$ on Cu(poly) & \cite{Shuai13} & 291.30 & 6.30 & & \\
	Monolayer physisorbed CO$_2$ on Cu(211) & \cite{Copperthwaite88} & 291.50 & 6.50 & & \\
	Physisorbed CO$_2$ on Cu(100) & \cite{Browne91} & 292.00 & 7.00 & & \\
	\hline		
	\end{tabular}
    
\textsuperscript{\emph{a}} All energies are given in eV.
\end{table*}

The calculated C1s binding energy shifts are compared against the available experimental data in Figure~\ref{Fig_C1s_shifts} and Table~\ref{Tab_C_ads}. In our calculations, adsorbed ethene is used as a model of ``adventitious carbon", and we use the C1s binding energy of this species as the reference for all theoretical C1s binding energy shifts. Experimental C1s binding energy shifts are calculated relative to a reference energy of 285.0~eV. In Figure~\ref{Fig_C1s_shifts} and Table~\ref{Tab_C_ads}, we have grouped together all peak assignments that are referred to as ``adventitious carbon", ``carbon contamination", ``graphitic carbon", ``C$^0$" or ``C$_x$H$_y$" in the experimental literature. The corresponding experimental binding energy shifts range from -0.6~eV to 0.2~eV \cite{Mudiyanselage13,Deng08,Copperthwaite88,Browne91,Favaro17}. 

For CO on Cu(111), the theoretical binding energy shift obtained for the 3-fold adsorption site (1.40~eV) is slightly closer to the experimental values reported for the (111) surface (1.1-1.2~eV) \cite{Eren15,Mudiyanselage13} than the theoretical value for the top site (1.75~eV). However, as discussed before, further work is required to assess the influence of the choice of exchange-correlation functional in the $\Delta$-SCF calculation on the core-level binding energy of CO on Cu(111). For the formate species, the theoretical binding energy of 2.81~eV agrees well with the majority of the published experimental values for formate on various Cu surfaces that range from 2.3~eV to 3.2~eV \cite{Nakamura97,Nishimura00,Deng08,Carley96,Pohl98,Favaro17}. The outlier amongst the experimental datapoints (at 4.75~eV \cite{Yang10}) is also the one that lies furthest from the calculated value. For methoxy on Cu(111), we note that unfortunately neither of the detailed photoelectron diffraction studies of this species \cite{Hofmann94,Carvalho92} report the experimental C1s binding energy. The binding energy shift that has been reported for C-O(H) environments on Cu(111) (1.3 eV) is relatively similar to our calculated value for methoxy on Cu(111) (1.82~eV), whereas the shifts that have been reported for the methoxy species on Cu(110) and polycrystalline Cu (0.8~eV and 0.2~eV \cite{Deng08,Carley96}) are much smaller. However, we believe that all of these experimental values should be taken with a note of caution, because they come from studies where complex surface chemical processes were investigated \cite{Carley96,Deng08,Favaro17}, making the interpretation of the experimental spectra extremely challenging. 

For physisorbed CO$_2$ on Cu(111), we have obtained a theoretical C1s binding energy shift of 4.92~eV. Favaro et al. \cite{Favaro17} have reported a binding energy shift of 3.4~eV for CO$_2$ on Cu(111), but the binding energy shifts reported for physisorbed CO$_2$ on other Cu surfaces and polycrystalline copper are significantly larger and range from 6.0~eV to 7.0~eV \cite{Copperthwaite88,Shuai13,Browne91}. Similarly large shifts of 6.2~eV and 6.5~eV have been reported for physisorbed CO$_2$ on Ni(110) and polycrystalline iron \cite{Illing88}, respectively. This suggests that the theoretical binding energy shift value of 4.92~eV is probably too low by approximately 1-2~eV. We note that in the calculations of free molecules, the C1s binding energy shift in CO$_2$ is also underestimated by $\sim$ 0.5~eV when using the PBE functional, but this is not sufficient to explain the discrepancy of more than 1~eV for the adsorbed species. In order to account for the remaining part of the disagreement, we hypothesize that CO$_2$ molecules may not physisorb onto Cu as a uniform monolayer. In particular, since the adsorption energy for CO$_2$ on Cu ($\sim$ 24~kJ/mol \cite{Muttaqien17_2}) is similar to the enthalpy of sublimation of solid CO$_2$ ($\sim$ 26~kJ/mol \cite{Chickos02}), the formation of three-dimensional adsorbed clusters may be favourable even at monolayer or sub-monolayer coverage. For CO$_2$ molecules in adsorbed clusters, it is reasonable to expect that the O1s binding energy is higher than for a single adsorbed molecule because the screening of the core hole in the final state is weaker when the molecule is located further away from the metal surface.

Finally, we note that we have not been able to calculate theoretical photoelectron binding energies for chemisorbed CO$_2$. In agreement with previous experimental \cite{Solymosi81,Graciani14,Solymosi91,Muttaqien17_2} and theoretical \cite{Wang07,Muttaqien17_2} investigations we have found that CO$_2$ does not chemisorb on defect-free Cu(111), which is the surface that has been considered throughout this work. In fact, the theoretical calculations of Muttaqien et al. suggest that the chemisorption of CO$_2$ is also unfavourable on ideal stepped and kinked Cu surfaces that expose the Cu(111) face \cite{Muttaqien14}, and very recently it has been proposed that the chemisorbed state can only occur on Cu(111) in the presence of sub-surface oxygen atoms \cite{Favaro17}.

\section{Conclusions}
In this work, we have shown that accurate core-level binding energy shifts of various adsorbed molecules on Cu(111) can be obtained from $\Delta$-SCF calculations on a cluster model of the surface. For the majority of the studied adsorbates (H$_2$O, OH, HCOO, C$_2$H$_4$ and $CO$), the calculated binding energy shifts agree well with published experimental data. In the few cases where the agreement is less good (CH$_3$O, CO$_2$), the theoretical results may indicate the need to re-examine experimental peak assignments. 

The ability to calculate core-level binding energy shifts from first principles is highly desirable due to the commonly encountered difficulties in the interpretation of experimental spectra. In particular, theoretical modelling may be the only practical way for estimating core-level binding energies of atoms in complex chemical environments, such as those that are formed under non-UHV conditions, under irradiation or in electrochemical setups.

On the other hand, the results presented in this work also highlight the difficulty of assessing the accuracy of a method for calculating XPS binding energy shifts for surface species which stems from the scarcity of reliable reference data. To make progress in this direction, it would be highly desirable to combine XPS measurements with other experimental techniques, such as scanning tunnelling microscopy or surface X-ray diffraction, to establish detailed structural models for several adsorbed species which could then inform first-principles calculations of core-level binding energies.

\section{Acknowledgements}
J.M.K. and J.L. acknowledge support from EPRSC under Grant No. EP/R002010/1 and also from the Thomas Young Centre under Grant No. TYC-101. Via J.L.'s membership of the UK's HEC
Materials Chemistry Consortium, which is funded by EPSRC (EP/L000202), this work used the ARCHER UK National Supercomputing Service.

\bibliography{reflib}

\begin{thebibliography}{10}

\bibitem{Newsome80}
D.~Newsome, ``The water-gas shift reaction,'' {\em Catalysis Reviews Science
  and Engineering}, vol.~21, p.~275, 1980.

\bibitem{Zhang17}
Z.~Zhang, S.-S. Wang, R.~Song, T.~Cao, L.~Luo, X.~Chen, Y.~Gao, J.~Lu, W.-X.
  Li, and W.~Huang, ``The most active cu facet for low-temperature water gas
  shift reaction,'' {\em Nature Communications, Volume 8, id. 488}, vol.~8,
  p.~488, sep 2017.

\bibitem{Nielsen18}
D.~Nielsen, X.-M. Hu, K.~Daasbjerg, and T.~Skrydstrup, ``Chemically and
  electrochemically catalysed conversion of co2 to co with follow-up
  utilization to value-added chemicals,'' {\em Nature Catalysis}, vol.~1,
  p.~244, 2018.

\bibitem{Kattel16}
S.~Kattel, B.~Yan, Y.~Yang, J.~G. Chen, and P.~Liu, ``Optimizing binding
  energies of key intermediates for co2 hydrogenation to methanol over
  oxide-supported copper.,'' {\em J. Am. Chem. Soc.}, vol.~138, no.~38,
  pp.~12440--50, 2016.

\bibitem{Behrens14}
M.~Behrens, ``Heterogeneous catalysis of co* conversion to methanol on copper
  surfaces.,'' {\em Angew. Chem. Int. Ed. Engl.}, vol.~53, no.~45,
  pp.~12022--4, 2014.

\bibitem{Meunier11}
F.~C. Meunier, ``Mixing copper nanoparticles and zno nanocrystals: a route
  towards understanding the hydrogenation of co2 to methanol?,'' {\em Angew.
  Chem. Int. Ed. Engl.}, vol.~50, no.~18, pp.~4053--4, 2011.

\bibitem{Eren15}
B.~Eren, C.~Heine, H.~Bluhm, G.~A. Somorjai, and M.~Salmeron, ``Catalyst
  chemical state during co oxidation reaction on cu(111) studied with
  ambient-pressure x-ray photoelectron spectroscopy and near edge x-ray
  adsorption fine structure spectroscopy.,'' {\em J. Am. Chem. Soc.}, vol.~137,
  no.~34, pp.~11186--90, 2015.

\bibitem{Deng08}
X.~Deng, A.~Verdaguer, T.~Herranz, C.~Weis, H.~Bluhm, and M.~Salmeron,
  ``Surface chemistry of cu in the presence of co2 and h2o,'' {\em Langmuir},
  vol.~24, p.~9474, 2008.

\bibitem{Mudiyanselage13}
K.~Mudiyanselage, S.~D. Senanayake, L.~Feria, S.~Kundu, A.~E. Baber,
  J.~Graciani, A.~B. Vidal, S.~Agnoli, J.~Evans, R.~Chang, S.~Axnanda, Z.~Liu,
  J.~F. Sanz, P.~Liu, J.~A. Rodriguez, and D.~J. Stacchiola, ``Importance of
  the metal-oxide interface in catalysis: in situ studies of the water-gas
  shift reaction by ambient-pressure x-ray photoelectron spectroscopy.,'' {\em
  Angew. Chem. Int. Ed. Engl.}, vol.~52, no.~19, pp.~5101--5, 2013.

\bibitem{Graciani14}
J.~Graciani, K.~Mudiyanselage, F.~Xu, A.~E. Baber, J.~Evans, S.~D. Senanayake,
  D.~J. Stacchiola, P.~Liu, J.~Hrbek, J.~F. Sanz, and J.~A. Rodriguez, ``Highly
  active copper-ceria and copper-ceria-titania catalysts for methanol synthesis
  from co2,'' {\em Science, Volume 345, Issue 6196, pp. 546-550 (2014).},
  vol.~345, pp.~546--550, aug 2014.

\bibitem{Rodriguez07}
J.~A. Rodriguez, P.~Liu, J.~Hrbek, J.~Evans, and M.~P{\'{e}}rez, ``Water gas
  shift reaction on cu and au nanoparticles supported on ceo2(111) and
  zno(0001): intrinsic activity and importance of support interactions.,'' {\em
  Angew. Chem. Int. Ed. Engl.}, vol.~46, no.~8, pp.~1329--32, 2007.

\bibitem{Favaro17}
M.~Favaro, H.~Xiao, T.~Cheng, W.~Goddard~III, J.~Yano, and E.~Crumlin,
  ``Subsurface oxide plays a critical role in co2 activation by cu(111)
  surfaces to form chemisorbed co2, the first step in reduction of co2,'' {\em
  PNAS}, vol.~114, p.~6706, 2017.

\bibitem{Nakamura97}
J.~Nakamura, Y.~Kushida, Y.~Choi, and T.~Uchijima, ``X-ray photoelectron
  spectroscopy and scanning tunnel microscope studies of formate species
  synthesized on cu(111) surfaces,'' {\em Journal of Vacuum Science and
  Technology A: Vacuum, Surfaces, and Films}, vol.~15, p.~1568, 1997.

\bibitem{Nishimura00}
H.~Nishimura, T.~Yatsu, T.~Fujitani, T.~Uchijima, and J.~Nakamura, ``Synthesis
  and decomposition of formate on a cu(111) surface {---} kinetic analysis,''
  {\em Journal of Molecular Catalysis A: Chemical}, vol.~155, p.~3, 2000.

\bibitem{Pohl98}
M.~Pohl and A.~Otto, ``Adsorption and reaction of carbon dioxide on pure and
  alkali-metal promoted cold-deposited copper films,'' {\em Surface Science,
  Volume 406, Issue 1, p. 125-137.}, vol.~406, pp.~125--137, may 1998.

\bibitem{Stacchiola15}
D.~J. Stacchiola, ``Tuning the properties of copper-based catalysts based on
  molecular in situ studies of model systems.,'' {\em Acc. Chem. Res.},
  vol.~48, no.~7, pp.~2151--8, 2015.

\bibitem{Carley94}
A.~Carley, M.~Roberts, and A.~Strutt, ``Chemical reactivity of co and co2 at a
  cu(110)-cs surface,'' {\em Catalysis Letters}, vol.~29, p.~169, 1994.

\bibitem{Shuai13}
W.~Shuai and X.~Guoqin, ``Effect of photon irradiation on the adsorption of co2
  on polycrystalline cu,'' {\em Chinese Journal of Catalysis}, vol.~34, p.~865,
  2013.

\bibitem{Copperthwaite88}
R.~G. Copperthwaite, P.~R. Davies, M.~A. Morris, M.~W. Roberts, and R.~A.
  Ryder, ``The reactive chemisorption of carbon dioxide at magnesium and copper
  surfaces at low temperature,'' {\em Catalysis Letters}, vol.~1, pp.~11--19,
  1988.

\bibitem{Browne91}
V.~M. Browne, A.~F. Carley, R.~G. Copperthwaite, P.~R. Davies, E.~M. Moser, and
  M.~W. Roberts, ``Activation of carbon dioxide at bismuth, gold and copper
  surfaces,'' {\em Applied Surface Science, Volume 47, Issue 4, p. 375-379.},
  vol.~47, pp.~375--379, jun 1991.

\bibitem{Carley96}
A.~Carley, A.~Chambers, P.~Davies, G.~Mariotti, R.~Kurian, and M.~Roberts,
  ``Surface oxygen and chemical specificity at copper and caesium surfaces,''
  {\em Faraday Discussions}, vol.~105, p.~225, 1996.

\bibitem{Minar2011}
J.~Minar, J.~Braun, S.~Mankovsky, and H.~Ebert, ``Calculation of angle-resolved
  photo emission spectra within the one-step model of photo emission - recent
  developments,'' {\em Journal of Electron Spectroscopy and Related Phenomena},
  vol.~184, p.~91, 2011.

\bibitem{Haverkort08}
M.~W. Haverkort, I.~S. Elfimov, L.~H. Tjeng, G.~A. Sawatzky, and A.~Damascelli,
  ``Strong spin-orbit coupling effects on the fermi surface of sr2ruo4 and
  sr2rho4,'' {\em Physical Review Letters, vol. 101, Issue 2, id. 026406},
  vol.~101, p.~026406, jul 2008.

\bibitem{Jin13}
W.~Jin, P.-C. Yeh, N.~Zaki, D.~Zhang, J.~T. Sadowski, A.~Al-Mahboob, A.~M.
  van~der Zande, D.~A. Chenet, J.~I. Dadap, I.~P. Herman, P.~Sutter, J.~Hone,
  and J.~Osgood, Richard~M., ``Direct measurement of the thickness-dependent
  electronic band structure of mos2 using angle-resolved photoemission
  spectroscopy,'' {\em Physical Review Letters, vol. 111, Issue 10, id.
  106801}, vol.~111, p.~106801, sep 2013.

\bibitem{Payne11}
D.~J. Payne, M.~D.~M. Robinson, R.~G. Egdell, A.~Walsh, J.~McNulty, K.~E.
  Smith, and L.~F.~J. Piper, ``The nature of electron lone pairs in bivo4,''
  {\em Applied Physics Letters, Volume 98, Issue 21, id. 212110 (3 pages)
  (2011).}, vol.~98, p.~212110, may 2011.

\bibitem{Kahk14}
J.~M. Kahk, C.~G. Poll, F.~E. Oropeza, J.~M. Ablett, D.~C{\'{e}}olin, J.-P.
  Rueff, S.~Agrestini, Y.~Utsumi, K.~D. Tsuei, Y.~F. Liao, F.~Borgatti,
  G.~Panaccione, A.~Regoutz, R.~G. Egdell, B.~J. Morgan, D.~O. Scanlon, and
  D.~J. Payne, ``Understanding the electronic structure of iro2 using
  hard-x-ray photoelectron spectroscopy and density-functional theory,'' {\em
  Physical Review Letters, Volume 112, Issue 11, id.117601}, vol.~112,
  p.~117601, mar 2014.

\bibitem{Lischner15}
J.~Lischner, G.~K. P{\'{a}}lsson, D.~Vigil-Fowler, S.~Nemsak, J.~Avila, M.~C.
  Asensio, C.~S. Fadley, and S.~G. Louie, ``Satellite band structure in silicon
  caused by electron-plasmon coupling,'' {\em Physical Review B, Volume 91,
  Issue 20, id.205113}, vol.~91, p.~205113, may 2015.

\bibitem{Bowker80}
M.~Bowker and R.~Madix, ``Xps, ups and thermal desorption studies of alcohol
  adsorption on cu(110),'' {\em Surface Science}, vol.~95, p.~190, 1980.

\bibitem{Yang10}
Y.~Yang, J.~Evans, J.~A. Rodriguez, M.~G. White, and P.~Liu, ``Fundamental
  studies of methanol synthesis from co(2) hydrogenation on cu(111), cu
  clusters, and cu/zno(0001).,'' {\em Phys Chem Chem Phys}, vol.~12, no.~33,
  pp.~9909--17, 2010.

\bibitem{Hofmann94}
P.~Hofmann, K.-M. Schindler, S.~Bao, V.~Fritzsche, D.~E. Ricken, A.~M.
  Bradshaw, and D.~P. Woodruff, ``The geometric structure of the surface
  methoxy species on cu(111),'' {\em Surface Science, Volume 304, Issue 1, p.
  74-84.}, vol.~304, pp.~74--84, mar 1994.

\bibitem{Au80}
C.~T. Au and M.~W. Roberts, ``Photoelectron spectroscopic evidence for the
  activation of adsorbate bonds by chemisorbed oxygen,'' {\em Chemical Physics
  Letters, Volume 74, Issue 3, p. 472-474.}, vol.~74, pp.~472--474, sep 1980.

\bibitem{Roberts14}
M.~Roberts, ``Low energy pathways and precursor states in the catalytic
  oxidation of water and carbon dioxide at metal surfaces and comparisons with
  ammonia oxidation,'' {\em Catalysis Letters}, vol.~144, p.~767, 2014.

\bibitem{Zhao17}
J.~Zhao, F.~Gao, S.~P. Pujari, H.~Zuilhof, and A.~V. Teplyakov, ``Universal
  calibration of computationally predicted n 1s binding energies for
  interpretation of xps experimental measurements.,'' {\em Langmuir}, vol.~33,
  no.~41, pp.~10792--10799, 2017.

\bibitem{Delesma18}
F.~A. Delesma, M.~Van~den Bossche, H.~Gr{\"{o}}nbeck, P.~Calaminici, A.~M.
  K{\"{o}}ster, and L.~G.~M. Pettersson, ``A chemical view on x-ray
  photoelectron spectroscopy: the esca molecule and surface-to-bulk xps
  shifts.,'' {\em Chemphyschem}, vol.~19, no.~2, pp.~169--174, 2018.

\bibitem{Kunkel18}
P.~J. R. J. A.~R. Christian~Kunkel, Francesc~Vi{\~{n}}es and F.~Illas,
  ``Combining theory and experiment for multitechnique characterization of
  activated co2 on transition metal carbide (001) surfaces,'' {\em The Journal
  of Physical Chemistry C}, 2018.

\bibitem{Artyushkova17}
K.~Artyushkova, I.~Matanovic, B.~Halevi, and P.~Atanassov, ``Oxygen binding to
  active sites of fe-n-c orr electrocatalysts observed by ambient-pressure
  xps,'' {\em The Journal of Physical Chemistry C}, vol.~121, p.~2836, 2017.

\bibitem{Bellafont17}
N.~Bellafront, F.~Vines, W.~Hieringer, and F.~Illas, ``Predicting core level
  binding energies shifts: Suitability of the projector augmented wave approach
  as implemented in vasp,'' {\em Journal of Computational Chemistry}, vol.~28,
  p.~518, 2017.

\bibitem{Artyushkova13}
K.~Artyushkova, B.~Kiefer, B.~Halevi, A.~Knop-Gericke, R.~Schlogl, and
  P.~Atanassov, ``Density functional theory calculations of xps binding energy
  shift for nitrogen-containing graphene-like structures.,'' {\em Chem. Commun.
  (Camb.)}, vol.~49, no.~25, pp.~2539--41, 2013.

\bibitem{Aizawa05}
T.~Aizawa, S.~Suehara, S.~Hishita, S.~Otani, and M.~Arai, ``Surface core-level
  shift and electronic structure on transition-metal diboride (0001)
  surfaces,'' {\em Physical Review B, vol. 71, Issue 16, id. 165405}, vol.~71,
  p.~165405, apr 2005.

\bibitem{vanSetten18}
M.~J. van Setten, R.~Costa, F.~Vi{\~{n}}es, and F.~Illas, ``Assessing gw
  approaches for predicting core level binding energies.,'' {\em J Chem Theory
  Comput}, vol.~14, no.~2, pp.~877--883, 2018.

\bibitem{Cavigliasso99}
G.~Cavigliasso, {\em Application of Density Functional Theory to the
  Calculation of Molecular Core-Electron Binding Energies}.
\newblock PhD thesis, The University of British Columbia, Vancouver, Canada,
  July 1999.

\bibitem{Cavigliasso99-1}
G.~Cavigliasso and D.~P. Chong, ``Accurate density-functional calculation of
  core-electron binding energies by a total-energy difference approach,'' {\em
  Journal of Chemical Physics, Volume 111, Issue 21, pp. 9485-9492 (1999).},
  vol.~111, pp.~9485--9492, dec 1999.

\bibitem{Bellafont2016}
N.~Bellafont, G.~Alvarez~Saiz, F.~Vines, and F.~Illas, ``Performance of
  minnesota functionals on predicting core-level binding energies of molecules
  containing main-group elements,'' {\em Theor. Chem. Acc.}, vol.~135, p.~35,
  2016.

\bibitem{Zhang14}
Y.-J. Zhang, V.~Sethuraman, R.~Michalsky, and A.~Peterson, ``Competition
  between co2 reduction and h2 evolution on transition-metal
  electrocatalysts,'' {\em ACS Catalysis}, vol.~4, p.~3742, 2014.

\bibitem{Gajdoš05}
M.~Gajdo{\v{s}} and J.~Hafner, ``Co adsorption on cu(1 1 1) and cu(0 0 1)
  surfaces: Improving site preference in dft calculations,'' {\em Surface
  Science, Volume 590, Issue 2-3, p. 117-126.}, vol.~590, pp.~117--126, oct
  2005.

\bibitem{Wang07}
S.-G. Wang, X.-Y. Liao, D.-B. Cao, C.-F. Huo, Y.-W. Li, J.~Wang, and H.~Jiao,
  ``Factors controlling the interaction of co2 with transition metal
  surfaces,'' {\em Journal of Physical Chemistry C}, vol.~111, p.~16934, 2007.

\bibitem{Muttaqien17_2}
F.~Muttaqien, Y.~Hamamoto, I.~Hamada, K.~Inagaki, Y.~Shiozawa, K.~Mukai,
  T.~Koitaya, S.~Yoshimoto, J.~Yoshinobu, and Y.~Morikawa, ``Co2 adsorption on
  the copper surfaces: van der waals density functional and tpd studies,'' {\em
  The Journal of Chemical Physics, Volume 147, Issue 9, id.094702}, vol.~147,
  p.~094702, sep 2017.

\bibitem{Muttaqien17}
F.~Muttaqien, {\em Chemistry of CO2 Adsorption and Reaction on the Copper
  Surfaces}.
\newblock PhD thesis, Department of Precision Science and Technology, Osaka
  University, July 2017.

\bibitem{Watson00}
G.~W. Watson, R.~P.~K. Wells, D.~J. Willock, and G.~J. Hutchings, ``{$\pi$}
  adsorption of ethene on to the {111} surface of copper. a periodic ab initio
  study of the effect of k-point sampling on the energy, atomic and electronic
  structure,'' {\em Surface Science, Volume 459, Issue 1, p. 93-103.},
  vol.~459, pp.~93--103, jul 2000.

\bibitem{Phatak09}
A.~Phatak, W.~Delgass, F.~Ribeiro, and W.~Schneider, ``Density functional
  theory comparison of water dissociation steps on cu, au, ni, pd, and pt,''
  {\em Journal of Physical Chemistry C}, vol.~113, p.~7269, 2009.

\bibitem{Michaelides06}
A.~Michaelides, ``Density functional theory simulations of water metal
  interfaces: waltzing waters, a novel 2d ice phase, and more,'' {\em Applied
  Physics A, Volume 85, Issue 4, pp.415-425}, vol.~85, pp.~415--425, dec 2006.

\bibitem{Michaelides07}
A.~Michaelides and K.~Morgenstern, ``Ice nanoclusters at hydrophobic metal
  surfaces.,'' {\em Nat Mater}, vol.~6, no.~8, pp.~597--601, 2007.

\bibitem{Carrasco12}
J.~Carrasco, A.~Hodgson, and A.~Michaelides, ``A molecular perspective of water
  at metal interfaces,'' {\em Nature Materials, Volume 11, Issue 8, pp. 667-674
  (2012).}, vol.~11, pp.~667--674, aug 2012.

\bibitem{Greeley02}
J.~Greeley and M.~Mavrikakis, ``Methanol decomposition on cu(111): A dft
  study,'' {\em Journal of Catalysis}, vol.~208, p.~291, 2002.

\bibitem{Chen06}
W.-K. Chen, S.-H. Liu, M.-J. Cao, C.-H. Lu, Y.~Xu, and J.-Q. Li, ``Adsorption
  of methanol and methoxy on cu(111) surface: A first-principles periodic
  density functional theory study,'' {\em Chinese Journal of Chemistry},
  vol.~24, p.~872, 2006.

\bibitem{Lian16}
X.~Lian, P.~Xiao, S.-C. Yang, R.~Liu, and G.~Henkelman, ``Calculations of oxide
  formation on low-index cu surfaces,'' {\em The Journal of Chemical Physics,
  Volume 145, Issue 4, id.044711}, vol.~145, p.~044711, jul 2016.

\bibitem{Giazonni09}
P.~Giannozzi, S.~Baroni, N.~Bonini, M.~Calandra, R.~Car, C.~Cavazzoni,
  D.~Ceresoli, G.~L. Chiarotti, M.~Cococcioni, I.~Dabo, A.~Dal~Corso,
  S.~de~Gironcoli, S.~Fabris, G.~Fratesi, R.~Gebauer, U.~Gerstmann,
  C.~Gougoussis, A.~Kokalj, M.~Lazzeri, L.~Martin-Samos, N.~Marzari, F.~Mauri,
  R.~Mazzarello, S.~Paolini, A.~Pasquarello, L.~Paulatto, C.~Sbraccia,
  S.~Scandolo, G.~Sclauzero, A.~P. Seitsonen, A.~Smogunov, P.~Umari, and R.~M.
  Wentzcovitch, ``Quantum espresso: a modular and open-source software project
  for quantum simulations of materials,'' {\em Journal of Physics: Condensed
  Matter, Volume 21, Issue 39, article id. 395502, 19 pp. (2009).}, vol.~21,
  p.~395502, sep 2009.

\bibitem{Garrity14}
K.~Garrity, J.~Bennett, K.~Rabe, and D.~Vanderbilt, ``Pseudopotentials for
  high-throughput dft calculations,'' {\em Computational Materials Science},
  vol.~81, p.~446, 2014.

\bibitem{Bengtsson99}
L.~Bengtsson, ``Dipole correction for surface supercell calculations,'' {\em
  Physical Review B (Condensed Matter and Materials Physics), Volume 59, Issue
  19, May 15, 1999, pp.12301-12304}, vol.~59, pp.~12301--12304, may 1999.

\bibitem{Perdew96}
J.~P. Perdew, K.~Burke, and M.~Ernzerhof, ``Generalized gradient approximation
  made simple,'' {\em Physical Review Letters, Volume 77, Issue 18, October 28,
  1996, pp.3865-3868}, vol.~77, pp.~3865--3868, oct 1996.

\bibitem{Grimme06}
S.~Grimme, ``Semiempirical gga-type density functional constructed with a
  long-range dispersion correction.,'' {\em J Comput Chem}, vol.~27, no.~15,
  pp.~1787--99, 2006.

\bibitem{Valiev10}
M.~Valiev, E.~J. Bylaska, N.~Govind, K.~Kowalski, T.~P. Straatsma, H.~J.~J.
  Van~Dam, D.~Wang, J.~Nieplocha, E.~Apra, T.~L. Windus, and W.~A. de~Jong,
  ``Nwchem: A comprehensive and scalable open-source solution for large scale
  molecular simulations,'' {\em Computer Physics Communications, Volume 181,
  Issue 9, p. 1477-1489.}, vol.~181, pp.~1477--1489, sep 2010.

\bibitem{Stevens92}
W.~Stevens, M.~Krauss, H.~Basch, and P.~Jasien, ``Relativistic compact
  effective potentials and efficient, shared-exponent basis sets for the
  third-, fourth-, and fifth-row atoms,'' {\em Canadian Journal of Chemistry},
  vol.~70, p.~612, 1992.

\bibitem{Jensen14}
F.~Jensen, ``Unifying general and segmented contracted basis sets. segmented
  polarization consistent basis sets.,'' {\em J Chem Theory Comput}, vol.~10,
  no.~3, pp.~1074--85, 2014.

\bibitem{Jensen06}
F.~Jensen, ``The basis set convergence of spin-spin coupling constants
  calculated by density functional methods.,'' {\em J Chem Theory Comput},
  vol.~2, no.~5, pp.~1360--9, 2006.

\bibitem{Zhao08}
Y.~Zhano and D.~Truhlar, ``The m06 suite of density functionals for main group
  thermochemistry, thermochemical kinetics, noncovalent interactions, excited
  states, and transition elements: two new functionals and systematic testing
  of four m06-class functionals and 12 other functionals,'' {\em Theor Chem
  Account}, vol.~120, p.~215, 2008.

\bibitem{Smith90}
H.~G. Smith, R.~Berliner, J.~D. Jorgensen, M.~Nielsen, and J.~Trivisonno,
  ``Pressure effects on the martensitic transformation in metallic lithium,''
  {\em Physical Review B (Condensed Matter), Volume 41, Issue 2, January 15,
  1990, pp.1231-1234}, vol.~41, pp.~1231--1234, jan 1990.

\bibitem{Barrett47}
C.~S. Barrett, ``A low temperature transformation in lithium,'' {\em Physical
  Review, vol. 72, Issue 3, pp. 245-245}, vol.~72, pp.~245--245, aug 1947.

\bibitem{Dovesi14}
R.~Dovesi, R.~Orlando, A.~Erba, C.~Zicovich-Wilson, B.~Civalleri, S.~Casassa,
  L.~Maschio, M.~Ferrabone, M.~De~La~Pierre, P.~D'Arco, Y.~Noel, M.~Causa,
  M.~Rerat, and B.~Kirtman, ``Crystal14: A program for the ab initio
  investigation of crystalline solids,'' {\em International Journal of Quantum
  Chemistry}, vol.~114, p.~1287, 2014.

\bibitem{Raval88}
R.~Raval, S.~F. Parker, M.~E. Pemble, P.~Hollins, J.~Pritchard, and M.~A.
  Chesters, ``Ft-rairs, eels and leed studies of the adsorption of carbon
  monoxide on cu(111),'' {\em Surface Science, Volume 203, Issue 3, p.
  353-377.}, vol.~203, pp.~353--377, sep 1988.

\bibitem{Stroppa07}
A.~Stroppa, K.~Termentzidis, J.~Paier, G.~Kresse, and J.~Hafner, ``Co
  adsorption on metal surfaces: A hybrid functional study with plane-wave basis
  set,'' {\em Physical Review B, vol. 76, Issue 19, id. 195440}, vol.~76,
  p.~195440, nov 2007.

\bibitem{Ren09}
X.~Ren, P.~Rinke, and M.~Scheffler, ``Exploring the random phase approximation:
  Application to co adsorbed on cu(111),'' {\em Physical Review B, vol. 80,
  Issue 4, id. 045402}, vol.~80, p.~045402, jul 2009.

\bibitem{Lopez01}
N.~Lopez and J.~K. N{\o}rskov, ``Synergetic effects in co adsorption on cu-pd(1
  1 1) alloys,'' {\em Surface Science, Volume 477, Issue 1, p. 59-75.},
  vol.~477, pp.~59--75, apr 2001.

\bibitem{Gajdos04}
M.~Gajdos, A.~Eichler, and J.~Hafner, ``Co adsorption on close-packed
  transition and noble metal surfaces: trends from ab initio calculations,''
  {\em Journal of Physics: Condensed Matter, Volume 16, Issue 8, pp. 1141-1164
  (2004).}, vol.~16, pp.~1141--1164, mar 2004.

\bibitem{Illing88}
G.~Illing, D.~Heskett, E.~W. Plummer, H.-J. Freund, J.~Somers, T.~Lindner,
  A.~M. Bradshaw, U.~Buskotte, M.~Neumann, U.~Starke, K.~Heinz, P.~L.
  De~Andres, D.~Saldin, and J.~B. Pendry, ``Adsorption and reaction of co 2 on
  ni{110}: X-ray photoemission, near-edge x-ray absorption fine-structure and
  diffuse leed studies,'' {\em Surface Science, Volume 206, Issue 1, p. 1-19.},
  vol.~206, pp.~1--19, nov 1988.

\bibitem{Carvalho92}
A.~V. de~Carvalho, M.~C. Asensio, and D.~P. Woodruff, ``Determination of the
  orientation of methoxy on cu(111) using x-ray photoelectron diffraction,''
  {\em Surface Science, Volume 273, Issue 3, p. 381-384.}, vol.~273,
  pp.~381--384, jul 1992.

\bibitem{Chickos02}
J.~S. Chickos and J.~Acree, William~E., ``Enthalpies of sublimation of organic
  and organometallic compounds. 1910-2001,'' {\em Journal of Physical and
  Chemical Reference Data, Volume 31, Issue 2, p.537-698}, vol.~31,
  pp.~537--698, jun 2002.

\bibitem{Solymosi81}
F.~Solymosi and J.~Kiss, ``Adsorption and decomposition of hnco on cu(111)
  surface studied by auger electron, electron energy loss and thermal
  desorption spectroscopy,'' {\em Surface Science, Volume 104, Issue 1, p.
  181-198.}, vol.~104, pp.~181--198, mar 1981.

\bibitem{Solymosi91}
F.~Solymosi, ``The bonding, structure and reactions of co2 adsorbed on clean
  and promoted metal surfaces,'' {\em Journal of Molecular Catalysis}, vol.~65,
  p.~337, 1991.

\bibitem{Muttaqien14}
F.~Muttaqien, Y.~Hamamoto, K.~Inagaki, and Y.~Morikawa, ``Dissociative
  adsorption of co2 on flat, stepped, and kinked cu surfaces,'' {\em The
  Journal of Chemical Physics, Volume 141, Issue 3, id.034702}, vol.~141,
  p.~034702, jul 2014.

\end{thebibliography}
\bibliographystyle{ieeetr}

\end{document}